\documentclass[english,onecolumn]{IEEEtran}
\usepackage[T1]{fontenc}
\usepackage[latin9]{inputenc}
\usepackage{color}
\usepackage{verbatim}
\usepackage{float}
\usepackage{amsthm}
\usepackage{amsmath}
\usepackage{amssymb}
\usepackage{graphicx}
\usepackage{esint}

\makeatletter
\theoremstyle{plain}
\newtheorem{thm}{\protect\theoremname}
\theoremstyle{definition}
\newtheorem{defn}{\protect\definitionname}
\theoremstyle{plain}
\newtheorem{cor}{\protect\corollaryname}
\theoremstyle{plain}
\newtheorem{lem}{\protect\lemmaname}

\makeatother

\providecommand{\corollaryname}{Corollary}
\providecommand{\definitionname}{Definition}
\providecommand{\lemmaname}{Lemma}
\providecommand{\theoremname}{Theorem}

\begin{document}

\title{An Efficient Feedback Coding Scheme with Low Error Probability for
Discrete Memoryless Channels}

\author{Cheuk Ting Li and Abbas El Gamal\\
 Department of Electrical Engineering\\
 Stanford University\\
 Stanford, California, USA\\
 Email: ctli@stanford.edu, abbas@ee.stanford.edu
 \thanks{
 This work is partially supported by Air Force grant FA9550-10-1-0124.  
 The work of C. T. Li was partially supported by a Hong Kong Alumni Stanford Graduate Fellowship.
 This paper was presented in part at the IEEE International Symposium on Information Theory, Honolulu, USA, June 2014.
 This paper is published in IEEE Transactions on Information Theory (Volume: 61, Issue: 6, June 2015), available at http://ieeexplore.ieee.org/document/7098395/ .
 Copyright (c) 2014 IEEE. Personal use of this material is permitted.  However, permission to use this material for any other purposes must be obtained from the IEEE by sending a request to pubs-permissions@ieee.org.
 }}
\maketitle
\begin{abstract}
Existing fixed-length feedback communication schemes are either specialized to particular channels (Schalkwijk--Kailath, Horstein), or apply to general channels but either have high coding complexity (block feedback schemes) or are difficult to analyze (posterior matching). This paper introduces a new fixed-length feedback coding scheme which achieves the capacity for all discrete memoryless channels, has an error exponent that approaches the sphere packing bound as the rate approaches the capacity, and has $O(n\log n)$ coding complexity. These benefits are achieved by judiciously combining features from previous schemes with new randomization technique and encoding/decoding rule. These new features make the analysis of the error probability for the new scheme easier than for posterior matching. 
\end{abstract}

\section{Introduction\label{sec:Introduction}}

\IEEEPARstart{S}{hannon} showed that feedback does not increase the capacity of memoryless
point-to-point channels~\cite{Shannon1956}. Feedback, however, has
many benefits, including simplifying coding and improving
reliability. Early examples of feedback coding schemes that demonstrate these
benefits include the Horstein~\cite{Horstein1963}, Zigangirov~\cite{zigangirov1970},
and Burnashev~\cite{burnashev1988} schemes for the binary symmetric
channel; and the Schalkwijk--Kailath scheme for the Gaussian channel~\cite{Schalkwijk--Kailath1966,Schalkwijk1966}.
Schalkwijk and Kailath showed that the error probability for their
scheme decays doubly exponentially in the block length. It is known, however,
that the error exponent for symmetric discrete memoryless channels with feedback
cannot exceed the sphere packing bound~\cite{dobrushin1962}.
Nevertheless, the schemes in~\cite{zigangirov1970,burnashev1988}
can attain better error exponents than the best known achievable error
exponent without feedback. D'yachkov~\cite{dyachkov1975} proposed
a general scheme for any discrete memoryless channel. The coding complexity for his scheme, however, appears to be very high.

In addition to the traditional fixed-length setting in which the number
of channel uses is predetermined before transmission commences, there
has been work on variable-length schemes in which transmission continues
until the error probability is lower than a prescribed target. The optimal
error exponent for this setting was given explicitly by Burnashev~\cite{Burnashev1976}.
Recently, Shayevitz and Feder~\cite{shayevitz2007,shayevitz2008,shayevitz2011}
introduced the posterior matching scheme, which unifies and extends
the Schalkwijk--Kailath and the Horstein schemes to general memoryless
channels. While they were able to show that the scheme achieves the
capacity for most of these channels in the variable-length setting,
their analysis of the error probability provides a lower bound that
is applicable only for low rates. A more general analysis of error
probability for variable-length schemes, including posterior matching,
is given in a recent paper by Naghshvar, Javidi and Wigger~\cite{naghshvar2013}.
Note that our focus here is only on fixed-length coding schemes for which the optimal
error exponent is not known in general.

In this paper, which is a more detailed version of our recent conference paper~\cite{Li--El-Gamal2014}, we propose a new fixed-length feedback coding scheme
for memoryless channels, which (i) achieves the capacity for all discrete
memoryless channels (DMCs), (ii) achieves an error exponent that approaches
the sphere packing bound for high rates (up to $O(\left(I\left(X;Y\right)-R\right)^{3})$),
and (iii) has coding complexity of only $O(n\log n)$ for discrete
memoryless channels. Our scheme is motivated by the posterior matching scheme.
However, unlike posterior matching, we assume a discrete message space,
e.g., as in the Burnashev scheme, apply a random cyclic shift
to the message points in each transmission, and use a {\em maximal information gain coding} rule instead of the actual posterior probability to simplify the analysis of the probability of error. This simplicity of analysis, however,
does not come at the expense of increased coding complexity relative
to posterior matching.

The rest of the paper is organized as follows. 
In the next section, we describe our feedback coding scheme and explain in detail how it differs from posterior matching. 
In Section \ref{sec:Analysis-PE}, we show that our scheme
achieves the capacity of any DMC, establish a lower bound on its
error exponent, and compare this bound to the sphere packing bound and bounds for other schemes. In Section~\ref{sec:Algorithms},
we discuss the scheme's coding complexity. Details of the coding algorithm and its complexity analysis are given in~\cite{feedbackcoding2013}.
\smallskip

\noindent{\bf Remark 1}: Throughout this paper, we use nats instead of
bits and $\ln$ instead of $\log$ to avoid adding normalization constants.
We denote the cumulative distribution function (cdf), the probability
mass function (pmf), and the probability density function (pdf) for
a random variable $X$ by $F_{X}$, $p_{X}$, and $f_{X}$, respectively.
We denote the set of integers $\{a,a+1,\dots,b\}$ as $[a:b]$. The
uniform distribution over $[0,1]$ is denoted by $\mathrm{U}[0,1]$.
The fractional part of $x$ is written as $x\,\mathrm{mod}\,1$.

\section{New Feedback Coding Scheme}\label{sec:Circular-Posterior-Matching}

Our scheme is motivated and is most similar to the posterior matching scheme~\cite{shayevitz2011}. Hence we begin with a brief description of posterior matching and its limitations, which have led to the development of our scheme. 

Posterior matching is a recursive coding scheme that achieves the
capacity of memoryless channels. Consider a memoryless channel $F_{Y|X}(y|x)$
with causal noiseless feedback, i.e., the transmitted symbol $X_{i}$ at time
$i$ is a function of the message and past received symbols
$Y^{i-1}$. Fix a distribution $F_{X}$ on the input symbols. The
message is represented by a real number $\Theta \sim \mathrm{U}[0,1]$.
The transmitted symbol at time $i$ is $X_{i}=F_{X}^{-1}(W_i)$, $W_i=F_{\Theta|Y^{i-1}}(\Theta|Y^{i-1})$,
where $F_{\Theta|Y^{i-1}}$, the posterior cdf of $\Theta$ given
the received symbols $Y^{i-1}$, is described recursively by
\begin{align*}
F_{\Theta|Y^{0}}(\theta) & =\theta,\\
F_{\Theta|Y^{i}}(\theta|y^{i}) & =F_{W|Y}(F_{\Theta|Y^{i-1}}(\theta|y^{i-1}) \, | \, y_{i}).
\end{align*}
Here $F_{W|Y}$ is the
cdf of $W$ conditioned on $Y$ assuming $W\sim \mathrm{U}[0,1]$ and $X=F_{X}^{-1}(W)$. If $Y$ is discrete, let 
\[
p_{Y|W}(y|w)=p_{Y|X}(y|F_{X}^{-1}(w)).
\]
Then, 
\begin{align}
\label{eq:FWgY} F_{W|Y} & \left(w|y\right)=\frac{\int_{0}^{w}p_{Y|W}\left(y|w'\right)\,dw'}{\int_{0}^{1}p_{Y|W}\left(y|w'\right)\, dw'}. 
\end{align}
The expression for continuous $Y$ can be given similarly.

Note that the posterior cdf $F_{\Theta|Y^{i}}$,
which can be regarded as the state of the transmission, forms a Markov
chain. To analyze the error probability, we can study the transition
of this Markov chain. However, the posterior cdf is a complicated
object. The analysis can be greatly simplified if a simpler object
(e.g., the posterior probability of the transmitted message) can be
used instead. As far as we know, this is not feasible due to the asymmetry
of the scheme in $\Theta$, in the sense that the behavior of the transition
of $F_{\Theta|Y^{i}}$ depends on the transmitted value of $\Theta$. Indeed, Shayevitz and Feder~\cite{shayevitz2011}
needed to use iterated function system to study the transition of the entire
posterior distribution, giving a rather complicated analysis of
the error probability of posterior matching that is applicable only
for rates below a certain threshold. Furthermore, this asymmetry results in messages having different error probabilities, which makes the maximal probability of error for the scheme worse than its average.

Our feedback coding scheme eliminates the aforementioned asymmetry of posterior matching resulting in all messages having the same error probability.
As a result, we are able to greatly simplify the analysis of the error probability and obtain a bound on the error exponent for {\em all} rates.

\begin{figure*}
\begin{centering}
\includegraphics[scale=0.2]{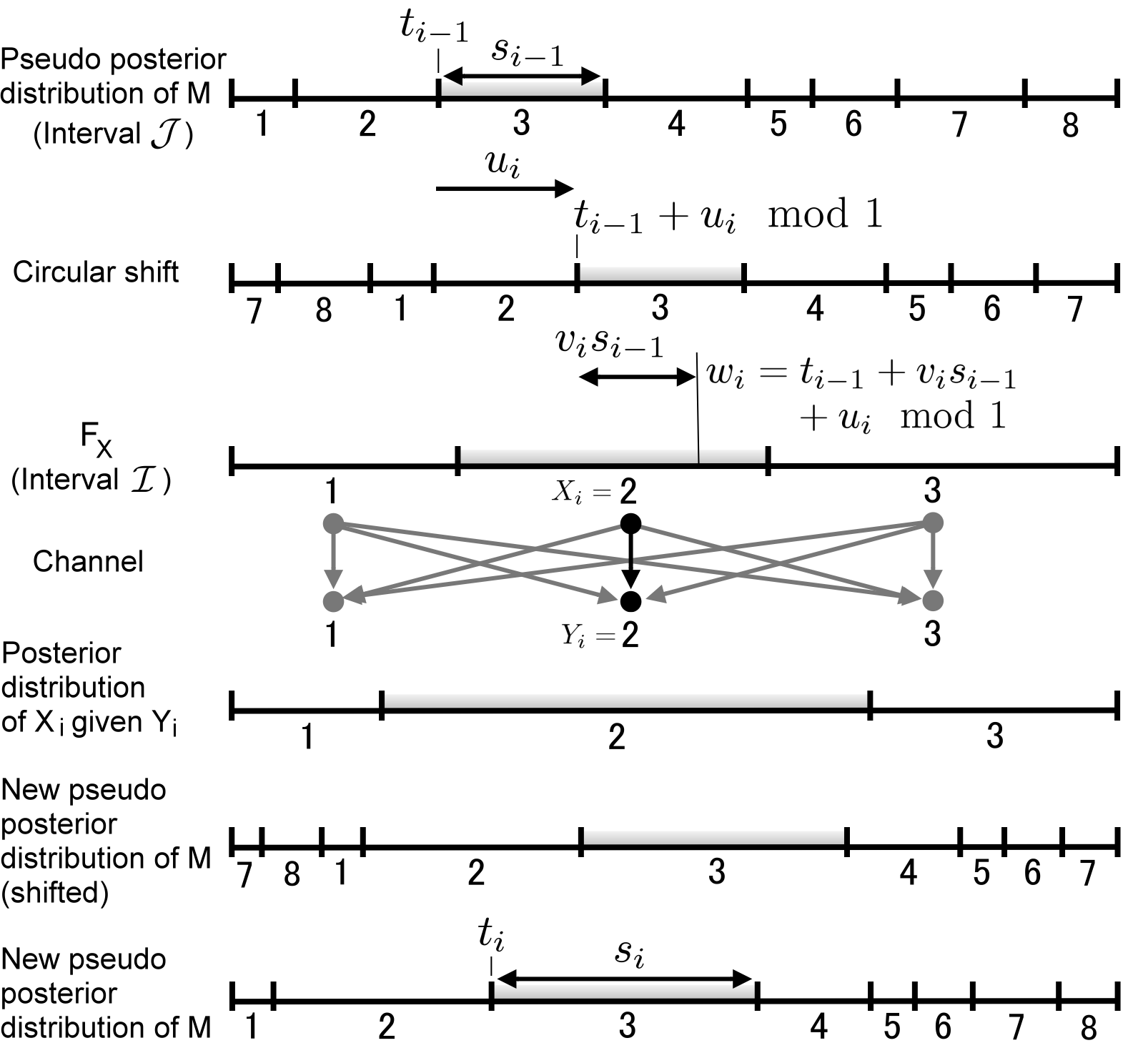} 
\par\end{centering}

\caption{\label{fig:illus_circ}Illustration of the new feedback scheme for
a DMC with input and output alphabet $\{1,2,3\}$. The message $M=3$
is transmitted. At time $i$, symbol $X_{i}=2$ is transmitted and
symbol $Y_{i}=2$ is received.}
\end{figure*}

Again consider a memoryless channel $F_{Y|X}(y|x)$ with causal noiseless
feedback. 
We describe our scheme with the aid of Figure~\ref{fig:illus_circ}. We assume that the message $M$ is uniformly distributed over $[1:e^{nR}]$ and 
represent message $m \in [1:e^{nR}]$ by the subinterval $[(m-1)e^{-nR},me^{-nR}]$ in $[0,1]$ (if the messages are not equally likely the subinterval length would be equal to the probability of the message). Fix the cdf $F_X(x)$ of the input symbol $X$ (which may be the capacity achieving distribution
for the channel), and partition the unit interval $\mathcal{I}$ according
to this distribution. The symbol to be transmitted at time $i$ is
determined as follows. The decoder, knowing $Y^{i-1}$, partitions
another unit interval $\mathcal{J}$ according to the pseudo posterior
probability distribution of $M$ given $Y^{i-1}$ (the details of
computing this distribution are described later). The encoder,
which has $Y^{i-1}$ via the feedback, also knows the partition of
$\mathcal{J}$. We denote the location of the left edge of the subinterval
corresponding to message $m$ by $t_{i-1}(m,y^{i-1},u^{i-1})$ (or $t_{i-1}(m)$ in short) and its length by $s_{i-1}(m,y^{i-1},u^{i-1})$ (or $s_{i-1}(m)$ in short) .
All subintervals are {\em cyclically shifted} by an amount $U_{i}\sim\mathrm{U}[0,1]$,
which is generated independently for each $i$ and is known to both
the encoder and the decoder. In practice, $U_i$ can be generated using a random seed communicated to both the encoder and the decoder via the forward or feedback channel.

A point $w_{i}$ is then selected in the subinterval corresponding
to the transmitted message $m$ according to $w_{i}=\bigl(v_{i}\cdot s_{i-1}(m)+t_{i-1}(m)\bigr)+u_{i}\,\,\mathrm{mod}\,1$,
where $v_{i}\in[0,1]$ is selected using a greedy rule to be described
later. The symbol to be transmitted at time $i$ is the one corresponding
to the subinterval in $\mathcal{I}$ which contains $w_{i}$. At the
end of communication, the decoder outputs the message $m$ corresponding
to the subinterval with the greatest length $s_{n}(m)$.

We are now ready to formally describe our scheme. At time $i\in[1:n]$,
the encoder transmits 
\[
X_{i}=F_{X}^{-1} (W_i), \quad W_i = w_{i}(M,Y^{i-1},U^{i},V_{i}),
\]
where{} 
\begin{align}
w_{i}(m,y^{i-1},u^{i},v_{i})\! & =\!v_{i}(m,y^{i-1},u^{i})\!\cdot\! s_{i-1}(m,y^{i-1},u^{i-1})\nonumber\\
&\;\;\,+t_{i-1}(m,y^{i-1},u^{i-1})+u_{i}\,\mathrm{mod}\,1,\nonumber \\
s_{0}\left(m\right) & =e^{-nR},\nonumber \\
t_{0}\left(m\right) & =\left(m-1\right)e^{-nR},\label{eq:circpost_recurrep}\\
s_{i}(m,y^{i},u^{i}) & =\int_{[t_{i-1}(m),t_{i-1}(m)+s_{i-1}(m)]+u_{i}\,\mathrm{mod}\,1}\!\!\!\!\!\!\!\!\!\!\!\!\!\!\!\!\!\!\!\!\!\!\!\!\!\!\!\!\!\!\!\!\!\!\!\!\!\!\!\!\!\!\!\!\!\!\!\!\!\!\!\!\!\!\!\!\!\!\!\!\!\!\!\!\!\!\!\!\!\!\!\!dF_{W|Y}(w|y_{i}),\nonumber\\
t_{i}(m,y^{i},u^{i}) & =\sum_{m'<m}s_{i}(m',y^{i},u^{i}),\nonumber 
\end{align}
where $F_{W|Y}$ is given in \eqref{eq:FWgY}. Note that in the above
integral we used the notation $[t,t+s]+u\mod1$ to mean the set $\{x+u\mod\,1\,:\, x\in[t,t+s]\}$.

Assuming message $m$ is transmitted, the encoder selects{}
$v_{i}(m,y^{i-1},u^{i})\in[0,1]$ using the \emph{maximal information
gain} rule 
\begin{align}
v_{i}(m,y^{i-1},u^{i})=&\arg\max_{v\in[0,1]}\mathbb{E}\Bigl[\ln s_{i}\left(m,(y^{i-1},Y_{i}),u^{i}\right)\nonumber\\
&\;\;\;\;\;\;\Bigl|W_{i}=w_{i}(m,y^{i-1},u^{i},v)\Bigr],\label{eq:maxinfogain}
\end{align}
where $Y_{i}$ is distributed according to $F_{Y|W}(y|w_{i})$. Note that this is a greedy rule that maximizes the ``information gain" for each
channel use.
\smallskip

We now provide explanations for the main ingredients of our scheme.

\noindent{1)} To explain the rule for selecting $X_{\ensuremath{i}}$ in \eqref{eq:circpost_recurrep},
note that at time $i$, both the encoder and the decoder know $Y^{i-1}$.
The encoder generates $X_{i}(M,Y^{i-1})$ that follows $F_{X}$
as closely as possible. For a DMC, 
\[
\mathbb{P}\{X_{i}=x\,|\, Y^{i-1}\!=\!y^{i-1}\}\!=\!\!\sum_{m\,:\, x_{i}(m,y^{i-1})=x}\!\!\!\!\!\!\!p_{M|Y^{i-1}}(m\,|\, y^{i-1}).
\]
Therefore, the distribution of $X_{i}$ is determined by how we divide
the posterior probabilities of the message among the input symbols.
If $M$ is continuous, we use the same trick as in posterior
matching, that is, $X_{i}=F_{X}^{-1}\circ F_{M|Y^{i-1}}(M\,|\, y^{i-1})$,
and $X_{i}$ would follow $F_{X}$. Since in our setting $M$ is discrete,
the posterior cdf $F_{M|Y^{i-1}}$ contains jumps, and each message
$m$ is mapped to an interval instead of a single point. We use $V_{i}$
to select a point on the interval and map it by $F_{X}^{-1}$ to obtain
the input symbol.
\smallskip

\noindent{2)} To explain the need for the circular shift of the
intervals via $U_{i}$, note that if we map a point on the interval
directly to the input symbol, the chosen symbol would depend on both
the position and the length of the interval corresponding to the correct
message. While the length of the interval provides information about
the posterior probability of the message, the position of the interval
does not contain any useful information. By applying the random circular
shift $U_{i}$, the analysis of the error probability involves only
the interval lengths. Suppose $m$ is sent, define $S_{i}=s_{i}(m,Y^{i},U^{i})$
to be the {\em pseudo} posterior probability of the transmitted message
(the length of the interval) at time $i$ and $T_{i}=t_{i}(m,Y^{i},U^{i})$
(the position of the interval). Note that $\{S_{i}\}$ forms a Markov
chain, and its transition can be specified by
\[
S_{i}=\int_{[0,S_{i-1}]+\tilde{U}_{i}\,\mathrm{mod}\,1} dF_{W|Y}(w|Y_{i}),
\]
where $Y_{i}\sim p_{Y|W}(\,\cdot\,|W_{i})$ is independent of $\tilde{U}^{i}$,
$S^{i-1}$ and $Y^{i-1}$, and
\begin{align*}
W_{i} & =V_{i}\cdot S_{i-1}+\tilde{U}_{i}\,\mathrm{mod}\,1,\\
V_{i} & =\arg\max_{v\in[0,1]}\mathbb{E}\left[\ln S_{i}\,|\, W_{i}=v\cdot S_{i-1}+\tilde{U}_{i}\,\mathrm{mod}\,1\right],
\end{align*}
Note that $\tilde{U}_{i}=T_{i-1}+U_{i}\sim\mathrm{U}[0,1]$ is independent
of $\tilde{U}^{i-1}$, $S^{i-1}$, and $Y^{i-1}$. 

As a result of the
random circular shift, the analysis of error reduces to studying the
real-valued Markov chain $\{S_{i}\}$. This is simpler than the analysis
of posterior matching, which involves keeping track of the entire
posterior distribution.

\smallskip

\noindent{3)} The reason we use the maximal information gain rule in~\eqref{eq:maxinfogain}
to select $V_{i}$ is that it yields a better bound on the error exponent
than the simpler rule of selecting $V_{i}$ uniformly at random. With
this complicated rule, however, it is very difficult to calculate
the posterior probabilities. Hence, in our scheme, the interval length
$s_{i}(m,y^{i},u^{i})$ is an {\em estimate} of the posterior probability
assuming $V_{i}$ is selected uniformly at random. In the following
we explain the method of estimating the posterior probability in detail.

Define another probability distribution $\tilde{\mathbb{P}}$ on $(M,X^{n},Y^{n},W^{n},U^{n},V^{n})$
in which $X^{n}$ is also generated according to~\eqref{eq:circpost_recurrep}
but $V^{n}$ is an i.i.d. sequence with $V_{i}\sim\mathrm{U}[0,1]$
instead of using~\eqref{eq:maxinfogain}. The receiver uses this
distribution to estimate the posterior probability of
each message, i.e., 
\[
s_{i}(m,y^{i},u^{i})=\tilde{\mathbb{P}}\{M=m\,|\, Y^{i}=y^{i},\, U^{i}=u^{i}\}.
\]
The expression in~\eqref{eq:circpost_recurrep} is obtained inductively
using{\allowdisplaybreaks 
\begin{align*}
 \lefteqn{\tilde{\mathbb{P}}\{M=m\,|\, Y^{i}=y^{i},\, U^{i}=u^{i}\}}\;\;& \\
 & \propto\tilde{\mathbb{P}}\{M=m\,|\, Y^{i-1}=y^{i-1},\, U^{i}=u^{i}\} \\
 & \;\;\;\;\cdot\tilde{\mathbb{P}}\{Y_{i}=y_{i}\,|\, M=m,\, Y^{i-1}=y^{i-1},\, U^{i}=u^{i}\},
\end{align*}
} where {\allowdisplaybreaks 
\begin{align*}
 & \tilde{\mathbb{P}}\{M=m\,|\, Y^{i-1}=y^{i-1},\, U^{i}=u^{i}\}\\
 & \;\cdot\tilde{\mathbb{P}}\{Y_{i}=y_{i}\,|\, M=m,\, Y^{i-1}=y^{i-1},\, U^{i}=u^{i}\}\\
 & =s_{i-1}(m)\cdot\tilde{\mathbb{P}}\{Y_{i}=y_{i}\,|\, M=m, Y^{i-1}=y^{i-1}, U^{i}=u^{i}\}\\
 & =s_{i-1}(m)\cdot\int_{0}^{1}\tilde{\mathbb{P}}\{Y_{i}=y_{i}\,|\, M=m,\, Y^{i-1}=y^{i-1},\\
 & \;\;\;\;\;\; U^{i}=u^{i},\, V_{i}=v\}\, dv\\
 & =s_{i-1}(m)\cdot\int_{0}^{1}\tilde{\mathbb{P}}\{Y_{i}=y_{i}\,|\, M=m,\, Y^{i-1}=y^{i-1},\\
 & \;\;\;\;\;\; U^{i}=u^{i}, W_{i}=v\!\cdot\! s_{i-1}(m)+t_{i-1}(m)+u_{i}\,\mathrm{mod}\,1\} dv\\
 & =\int_{[t_{i-1}(m),t_{i-1}(m)+s_{i-1}(m)]+u_{i}\,\mathrm{mod}\,1} \!\!\!\!\!\!\!\!\!\!\!\!\!\!\!\!\!\!\!\!\!\!\!\!\!\!\!\!\!\!\!\!\!\!\!\!\!\!\!\!\!\!\!\!\!\!\!\!\!\!\!\!\!\!\!\!\!\!\!\!\!\!\!\!\!\!\!\!\!\!\!\!\!\!\!\!\!\!\tilde{\mathbb{P}}\{Y_{i}=y_{i}\,| M=m,  Y^{i-1}=y^{i-1}\!, U^{i}\!=u^{i}, W_{i}=w\} dw\\
 & =\int_{[t_{i-1}(m),t_{i-1}(m)+s_{i-1}(m)]+u_{i}\,\mathrm{mod}\,1} \!\!\!\!\!\!\!\!\!\!\!\!\!\!\!\!\!\!\!\!\!\!\!\!\!\!\!\!\!\!\!\!\!\!\!\!\!\!\!\!\!\!\!\!\!\!\!\!\!\!\!\!\!\!\!\!\!\!\!\!\!\!\!\!\!\!\!\!\!\!\!\!\! p_{Y|W}(y_{i}|w)\, dw.
\end{align*}
Note that we write $s_{i-1}(m)=s_{i-1}(m,y^{i-1},u^{i-1})$ and
$t_{i-1}(m)=t_{i-1}(m,y^{i-1},u^{i-1})$ for simplicity. Hence, {\allowdisplaybreaks
\begin{align*}
 & \tilde{\mathbb{P}}\{M=m\,|\, Y^{i}=y^{i},\, U^{i}=u^{i}\}\\
 & =\!\frac{\int_{[t_{i-1}(m),t_{i-1}(m)+s_{i-1}(m)]+u_{i}\,\mathrm{mod}\,1}p_{Y|W}(y_{i}|w)\, dw}{\sum_{\tilde{m}=1}^{|\mathcal{M}|}\!\int_{[t_{i-1}(\tilde{m}),t_{i-1}(\tilde{m})+s_{i-1}(\tilde{m})]+u_{i}\,\mathrm{mod}\,1}\! p_{Y|W}\!(y_{i}|w) dw}\\
 & =\frac{\int_{[t_{i-1}(m),t_{i-1}(m)+s_{i-1}(m)]+u_{i}\,\mathrm{mod}\,1}p_{Y|W}(y_{i}|w)\, dw}{\int_{0}^{1}p_{Y|W}\left(y_{i}|w\right)dw}\\
 & =\int_{[t_{i-1}(m),t_{i-1}(m)+s_{i-1}(m)]+u_{i}\,\mathrm{mod}\,1}\, dF_{W|Y}(w|y_{i}).
\end{align*}
}} The quantity $s_{i}(m,y^{i},u^{i})$ can be viewed as a \emph{pseudo
posterior probability} of message $m$. Note that the pseudo posterior
probabilities of all the messages still sum up to one, hence we know
the correct message is recovered when its pseudo posterior probability
is greater than $1/2$.

\smallskip
From the above description, the key differences between
our scheme and posterior matching are as follows:
\smallskip

\noindent{1)} We apply a random circular shift $U_{i}$ to reduce
the analysis of error to studying the behavior of the Markov chain
$\{S_{i}\}$.
\smallskip

\noindent{2)} The message is an integer $M\in[1:e^{nR}]$ rather
than a real number $\Theta\in[0,1]$. This again simplifies the analysis.

\noindent{3)} Instead of using the posterior probability of the
message as in posterior matching, we use the maximal information gain
rule, which is crucial to the analysis of the scheme. 

\smallskip
As a result of these differences, our scheme can
achieve good error exponent over the entire rate range using a simpler
error probability analysis.

\section{Analysis of the Probability of Error}\label{sec:Analysis-PE}

In this section, we analyze the rate and the error exponent of our
scheme for DMCs. Note that in this case, $W_{i}=w\in[0,1]$ is mapped
to $X_{i}=x=F_{X}^{-1}\left(w\right)$ if $F_{X}(x-1)<w\le F_{X}(x)$.
As we discussed in the previous section, the pseudo posterior probability
of the transmitted message $\left\{ S_{i}\right\} $ forms a Markov
chain. We obtain the bound on the error exponent by analyzing this Markov chain.

In our scheme, the decoder declares $\hat{m}=\arg\max_{m'}s_{n}(m',y^{n},u^{n})$.
Since the pseudo posterior probabilities of all the messages sum up
to one, if the pseudo posterior probability of the transmitted message
$S_{n}=s_{n}(m,Y^{n},U^{n})>1/2$, we can be sure that the message
is recovered correctly. Hence, the probability of error is upper bounded
as 
\begin{align*}
P_{e}^{(n)} &=\mathbb{P}\left\{ M\neq\arg\max_{m}s_{n}(m,Y^{n},U^{n})\right\} \\
& \le\mathbb{P}\left\{ S_{n}\le1/2\right\}.
\end{align*}

\noindent {\bf Remark 2}: An alternative approach would
be to use a threshold decoder~\cite{martinez2011random}, which decodes to the message with
posterior probability greater than a threshold $\gamma$. However,
this would introduce another error event when there is a message other
than the correct one with pseudo posterior probability greater than
$\gamma$. As a result, we cannot analyze the error probability by
studying $S_{n}$ only. Therefore we fix the threshold at $1/2$ to
simplify the analysis.\smallskip{}

To study how the error probability decays with $n$, we consider the
error exponent 
\[
E(R)=\limsup_{n\to\infty}-n^{-1}\ln P_{e}^{(n)}(R).
\]
We define the moment generating function of the ideal
increment of information (or \emph{ideal moment generating function}
in short) for DMC as 
\begin{align*}
\phi\left(\rho\right) & =\sum_{x}p(x)\sum_{y}p(y|x)\left(\frac{p(x|y)}{p(x)}\right)^{-\rho}\\
 & =\sum_{x}p(x)\sum_{y}p(y|x)\left(\frac{p(y|x)}{\sum_{x'}p(x')p(y|x')}\right)^{-\rho}.
\end{align*}
The function $\ln\phi\left(\rho\right)$ is convex, and it is not
difficult to check that 
\[
\phi'\left(0\right)=\left.\frac{d}{d\rho}\phi\left(\rho\right)\right|_{\rho=0}=\left.\frac{d}{d\rho}\ln\phi\left(\rho\right)\right|_{\rho=0}=-I(X;Y).
\]
Similarly, we define the moment generating function of the actual
increment of information at $s$ (or \emph{actual moment generating
function} in short) as 
\[
\psi_{s}\left(\rho\right)=\mathbb{E}\left[\left.S_{i}^{-\rho}/S_{i-1}^{-\rho}\right|S_{i-1}=s\right].
\]
%

The function $\ln\psi_{s}\left(\rho\right)$ is convex. To obtain
the bound on the error exponent, we also need the quantity 
\[
\Psi=\inf_{\tau\left(s\right)}\sup_{s\in(0,1)}\psi_{s}\left(\tau\left(s\right)\right),
\]
where $\tau(s)$ is nondecreasing and the infimum is taken over all
nondecreasing functions $\tau:(0,1)\to[0,\infty)$. We have $\Psi\le1$,
since we can take $\tau\left(s\right)=0$.

We introduce the following condition on a DMC, which is sufficient
for our scheme to achieve the capacity. 
\begin{defn}
A pair of input symbols $x_{1}\neq x_{2}$ in a DMC $p(y|x)$ is said
to be redundant if $p(y|x_{1})=p(y|x_{2})$ for all $y$. 
\end{defn}
Note that if the channel has redundant input symbols, we can always
use only one of these symbols and ignore the others. Therefore we
can assume without loss of generality that the channel has no redundant
input symbols.

We are now ready to state the main result of this paper. 
\begin{thm}
\label{thm:errexp_1pass} For any DMC $p(y|x)$ without redundant
input symbols, we have $\Psi<1$, and the maximal information gain
scheme can achieve the capacity. Further, for any $R<I\left(X;Y\right)$,
the error exponent is lower bound as 
\begin{align*}
E(R) & \ge\sup_{\rho>0}\left\{ -\rho R-\ln\max\left(\phi\left(\rho\right),\,\Psi\right)\right\} .
\end{align*}

\end{thm}
The proof of this theorem is detailed in the following subsection.
\smallskip

The bound on the error exponent of our scheme becomes quite tight
as the rate tends to the capacity.
\begin{cor}
\label{cor:ratetocapacity}The error exponent $E(R)$
satisfies
\[
E(R)=\frac{\left(I\left(X;Y\right)-R\right)^{2}}{2\mathrm{Var}[\ln(p(Y|X)/p(Y))]}-O\left(\left(I\left(X;Y\right)-R\right)^{3}\right)
\]
as $R$ tends to $I\left(X;Y\right)$.
\end{cor}
The quantity $\mathrm{Var}[\ln(p(Y|X)/p(Y))]$ is known as the channel
dispersion~\cite{strassen1962,polyanskiy2010}. Note that this is
the same limit as for the sphere packing bound. Hence the error exponent
of our scheme tends to the sphere packing bound when the rate tends
to the capacity. The proof of this corollary is given in Appendix~\ref{coro}. 
\smallskip

To illustrate the above results, consider the following.

\noindent{\bf Example} (Binary Symmetric Channel):

Consider a binary symmetric channel with crossover probability $p$. It is well known that the
capacity of this channel is achieved with $X\sim\mathrm{Bern}(1/2)$. The
maximal information gain rule always selects the input symbol whose
probability interval has the larger overlapping area with the message
interval. The actual moment generating function is
$\psi_{s}(\rho) = p \alpha + q \beta$,
where $q=1-p$, and

\[
\begin{aligned}
\alpha\! & =\!\begin{cases}
(2p)^{-\rho}\bigg(1-2s+\frac{4sp}{(\rho-1)(q-p)}\bigg)+\frac{2s}{\rho-1} & s\le\frac{1}{2},\\[6pt]
(2s-1)\big(2q-\frac{q-p}{s}\big)^{-\rho}-\frac{2s(1-(2q-(q-p)/s)^{1-\rho})}{(\rho-1)(q-p)}\!\! & s>\frac{1}{2},
\end{cases}\end{aligned}
\]

\[
\begin{aligned}
\beta\! & =\!\begin{cases}
(2q)^{-\rho}\bigg(1-2s-\frac{4sq}{(\rho-1)(q-p)}\bigg)+\frac{2s}{\rho-1} & s\le\frac{1}{2},\\[6pt]
(2s-1)\big(2p+\frac{q-p}{s}\big)^{-\rho}+\frac{2s(1-(2p-(q-p)/s)^{1-\rho})}{(\rho-1)(q-p)}\!\! & s>\frac{1}{2}.
\end{cases}\end{aligned}
\]

The value of $\Psi$ can be found approximately using
dynamic programming. For example, for $\mathrm{BSC}(0.1)$, $\Psi\approx0.8948$.

\begin{figure*}
\begin{centering}
\includegraphics[scale=0.5]{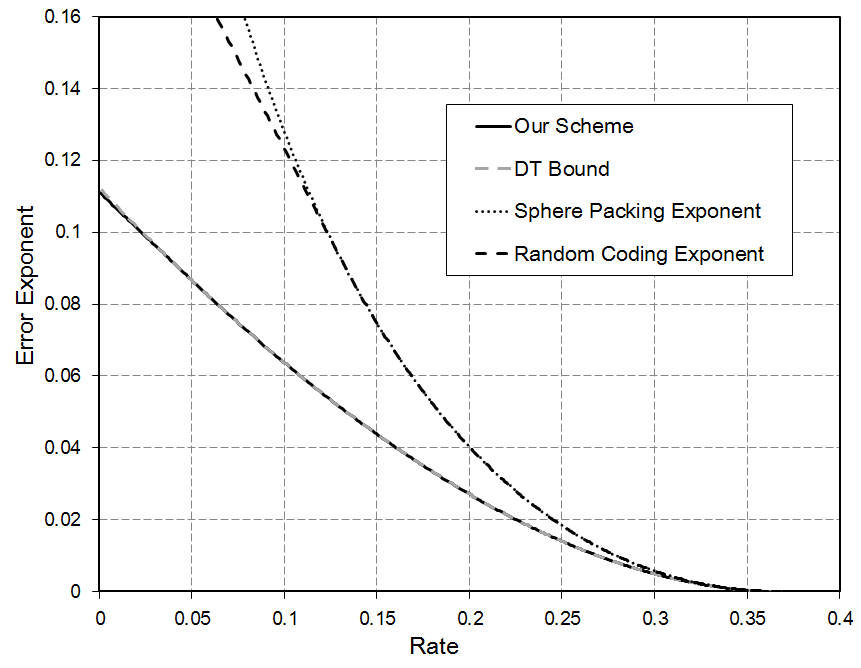} 
\par\end{centering}

\caption{\label{fig:errexp_bsc}Comparisons of the bound on the error error exponent for a $\mathrm{BSC}(0.1)$.}
\end{figure*}

Figure \ref{fig:errexp_bsc} compares the bound on the error exponent
for our scheme to the following.
\smallskip

\noindent{1)} The sphere packing exponent $\max_{Q}\max_{\rho>0}E_{0}\left(\rho,Q\right)-\rho R$,
where 
\[E_{0}\left(\rho,Q\right)=-\ln\sum_{y}\left(\sum_{x}Q\left(x\right)p\left(y|x\right)^{1/\left(1+\rho\right)}\right)^{1+\rho}.\]
\smallskip

\noindent{2)} The random coding exponent $\max_{Q}\max_{\rho\in(0,1]}E_{0}\left(\rho,Q\right)-\rho R$,
which is a lower bound without feedback.
\smallskip

\noindent{3)} The dependence-testing (DT) bound~\cite{martinez2011random},
which is the error exponent for random coding without feedback when
a threshold decoder is used.
\smallskip

Note that our error exponent approaches the sphere packing exponent
when $R$ is close to the capacity. Also our exponent
almost coincides with the DT bound, with noticeable difference only
when the rate is close to zero.

\subsection{Proof of Theorem~\ref{thm:errexp_1pass}} \label{sub:Proof-of-Theorem}

We first outline the main ideas of the proof of the
theorem. As we discussed, to analyze the error probability of our
scheme, it suffices to study how $S_{i}$ increases from $S_{0}=e^{-nR}$
to $S_{n}>1/2$. We divide the analysis of the scheme by the stage
of transmissions into: the \emph{starting phase}, 
where $S_{i}$ is small, the \emph{transition phase}, 
where $S_{i}$ is not close to $0$ or $1$, and the \emph{ending
phase,} where $S_{i}$ is close to $1$. We outline the proof for each phase.\medskip{}

\smallskip

\begin{figure}
\begin{centering}
\includegraphics[scale=0.16]{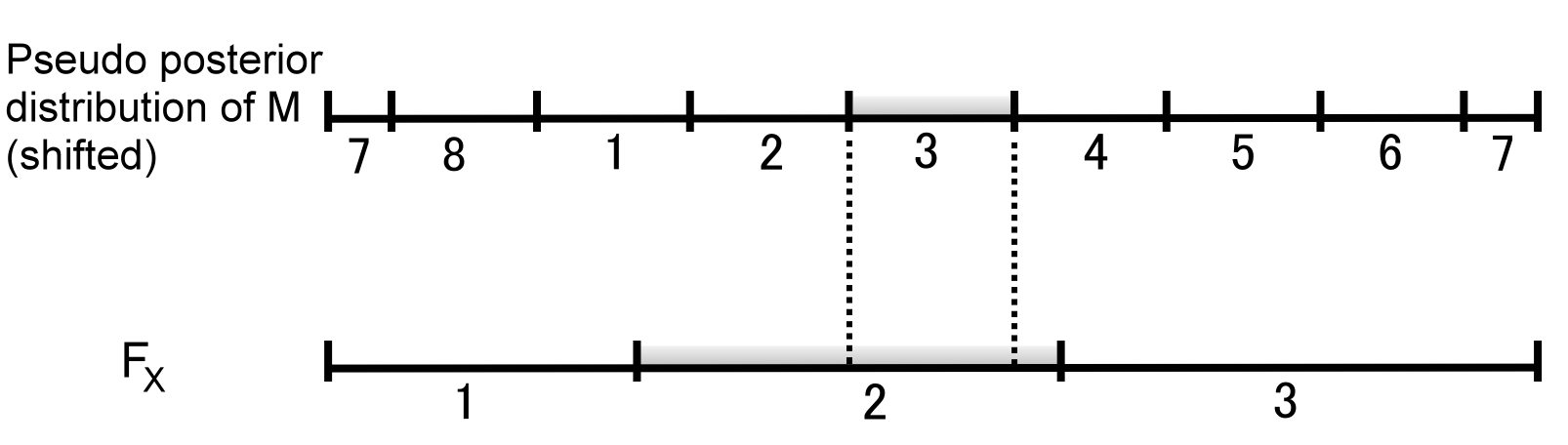} 
\par\end{centering}

\caption{\label{fig:phase1}Illustration of the shifted pseudo posterior distribution
during the starting phase (assuming $M=3$ is sent).}
\end{figure}

The starting phase refers to the transmission period in which
$S_{i}\le S_{\mathrm{start}}$, where $S_{\mathrm{start}}$ is a constant that depends on the channel. During this phase, the length
of the message interval $[T_{i-1}+U_{i},T_{i-1}+S_{i-1}+U_{i}]\mod1$
is close to $0$ and is very likely to overlap with the probability
interval $[F_{X}\left(x-1\right),F_{X}\left(x\right)]$ for only a single
input symbol $x$ as illustrated in Figure \ref{fig:phase1}. In this case, the maximal information gain rule
selects $x$ and the probability of $X_{i}$ would be close
to $p(x)$. The following lemma shows that in this regime the actual
moment generating function is close to the ideal one.
\begin{lem}[starting phase MGF]
\label{lem:smallprob}For any DMC $p(y|x)$ with
input pmf $p(x)$, let $S_{\mathrm{start}}=\min_{x\,:\, p(x)>0}p\left(x\right)$,
then there exists $\omega\ge1$ such that 
\begin{align*}
&\left(1-\frac{s}{S_{\mathrm{start}}}\right)\phi\left(\rho\right)+\frac{s}{S_{\mathrm{start}}}\omega^{-\rho}\le\psi_{s}\left(\rho\right) \\
&\qquad\qquad\qquad\qquad\qquad\le\left(1-\frac{s}{S_{\mathrm{start}}}\right)\phi\left(\rho\right)+\frac{s}{S_{\mathrm{start}}}\omega^{\rho}
\end{align*}
for $s\le S_{\mathrm{start}}$ and $\rho\ge0$.
\end{lem}
The proof of the lemma is given in Appendix~\ref{lem_start}.

\medskip{}

\smallskip

\begin{figure}
\begin{centering}
\includegraphics[scale=0.16]{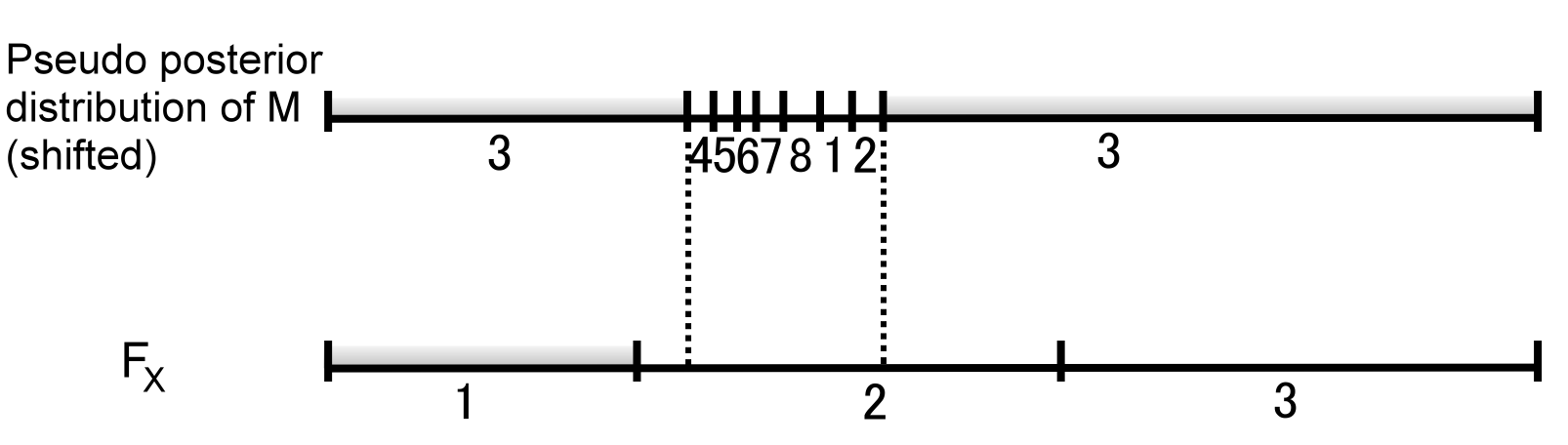} 
\par\end{centering}

\caption{\label{fig:phase3}Illustration of the shifted pseudo posterior distribution
during the ending phase (assuming $M=3$ is transmitted).}
\end{figure}

\begin{figure}
\begin{centering}
\includegraphics[scale=0.16]{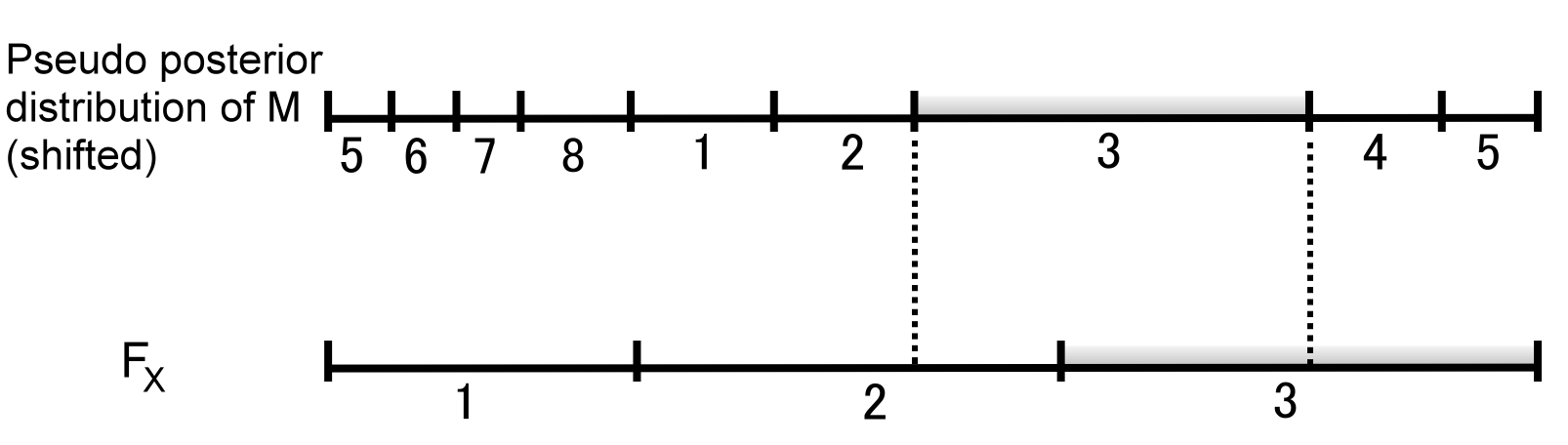} 
\par\end{centering}

\caption{\label{fig:phase2}Illustration of the shifted pseudo posterior distribution
during the transition phase (assuming $M=3$ is sent).}
\end{figure}

The ending phase refers to the transmission period in which $S_{i}\ge S_{\mathrm{end}}$,
where $S_{\mathrm{end}}$ is a constant that depends on the channel.
During the ending phase, the length of the message interval $[T_{i-1}+U_{i},T_{i-1}+S_{i-1}+U_{i}]\mod1$
is close to one. Hence, the maximal information gain rule is free
to select any input symbol. However, the complement of the message
interval is likely to overlap with only one symbol probability interval
$[F_{X}\left(\bar{x}-1\right),F_{X}\left(\bar{x}\right)]$ as illustrated in Figure \ref{fig:phase3}. In this
case, the maximal information gain rule selects the input symbol $x$, which is the
``opposite" of $\bar{x}$ in the sense that the posterior probability
of $X_{i}=\bar{x}$ is minimized when $X_{i}=x$ is transmitted. This would
maximize the posterior probability of the message. We can bound the
actual moment generating function during this phase as follows.

\medskip{}

\begin{lem}[ending phase MGF]
\label{lem:largeprob}For any DMC $p(y|x)$, there
exists $0<S_{\mathrm{end}}<1$, $\gamma>0$ and $\Psi_{\mathrm{end}}<1$
such that when $s\ge S_{\mathrm{end}}$, 
\[
\psi_{s}\left(\gamma\left(1-s\right)^{-1}\right)\le\Psi_{\mathrm{end}}.
\]

\end{lem}
The proof of the lemma is given in Appendix~\ref{lem_end}.

\begin{figure*}
\begin{centering}
\includegraphics[scale=0.21]{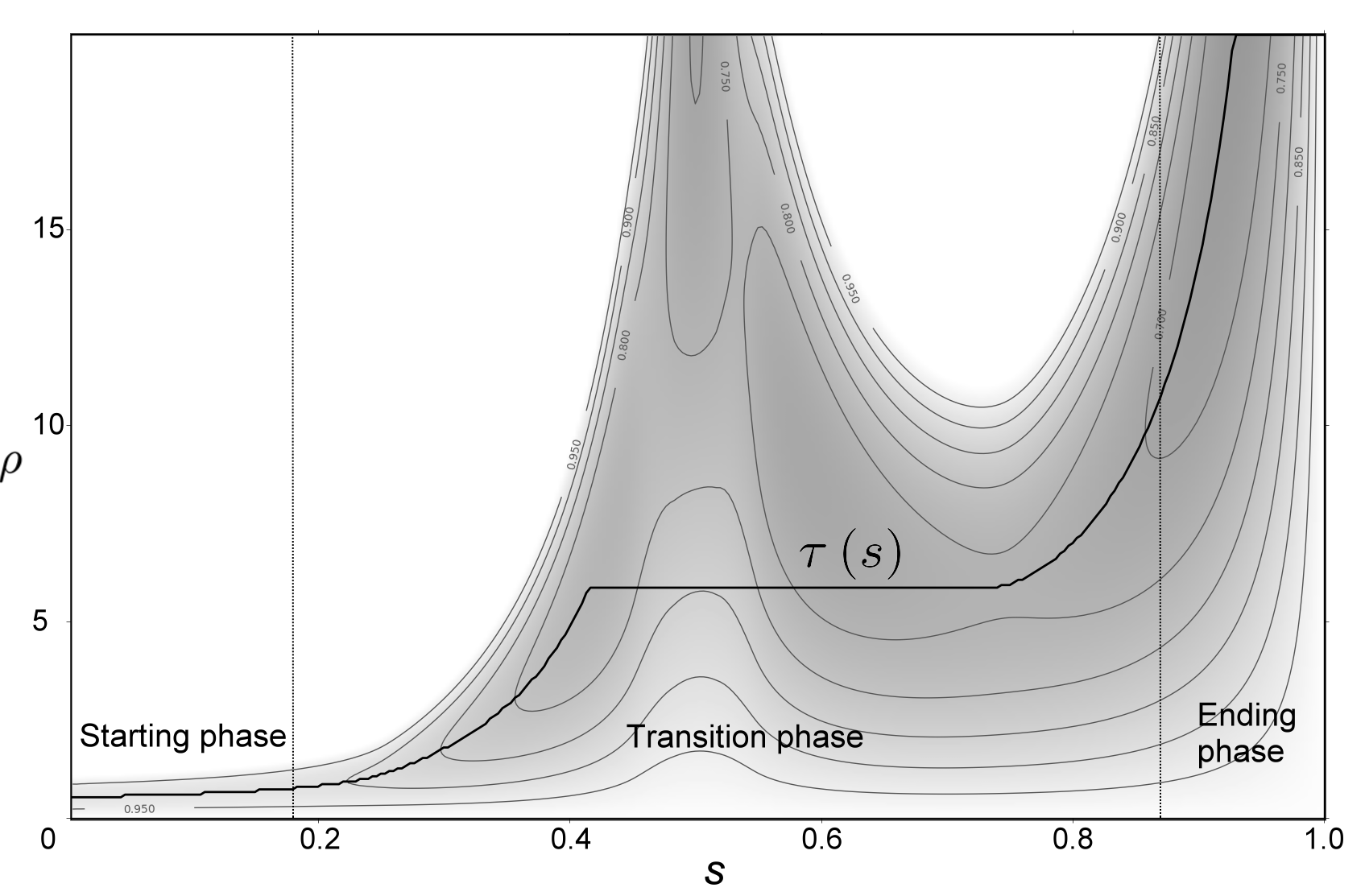} 
\par\end{centering}

\caption{\label{fig:illus_mgf}Contour plot of $\psi_{s}\left(\rho\right)$
for an example channel. Darker color indicates smaller $\psi_{s}\left(\rho\right)$.
The minimizing function $\tau\left(s\right)$ is also plotted.}
\end{figure*}

The transition phase refers to the transmission period in which $S_{\mathrm{start}}<S_{i}<S_{\mathrm{end}}$ as illustrated in Figure \ref{fig:phase2}.
For the
error exponent in Theorem \ref{thm:errexp_1pass} to be nonzero, we
need $\Psi<1$, therefore we need to find a nondecreasing function
$\tau:(0,1)\to[0,\infty)$ such that $\psi_{s}\left(\tau\left(s\right)\right)$
is bounded above and away from $1$. From the plot in Figure \ref{fig:illus_mgf},
we can see that $\psi_{s}\left(\rho\right)$ is well-behaved in the
starting and ending phases, but not in the transition phase. Nevertheless, it is
possible to construct $\tau$ satisfying the requirement, as shown in the following
lemma.

\begin{lem}
\label{lem:PsiLessThanOne} For a DMC $p(y|x)$ without
redundant input symbols, we have $\Psi<1$. 
\end{lem}
The proof of the lemma is given in Appendix~\ref{lem_trans}.

To show that our scheme achieves the capacity, recall
that $S_{i}$ should increase from $S_{0}=e^{-nR}$ to $S_{n}>1/2$,
or equivalently, $-\ln S_{i}$ should decrease from $-\ln S_{0}=nR$
to $-\ln S_{n}<\ln2$. When $n$ is large, $S_{0}$ is close to zero;
 hence the time spent in the starting phase would dominate. Since
the actual moment generating function is close to the ideal one during
this phase, we expect the decrease in $-\ln S_{i}$ for each
time step to be close to $-\phi'\left(0\right)=I(X;Y)$. Therefore
as long as $R<I(X;Y)$, $-\ln S_{i}$ would decrease from $nR$ to
a value smaller than $\ln2$ in $n$ time steps. However, we still
need to show that the transition and the ending phase would not affect
the performance of the code. As we will see in the proof of the theorem,
the fact that $\Psi<1$ is sufficient for this purpose. 

We now discuss the details of the proof of Theorem~\ref{thm:errexp_1pass}.
\smallskip

Let $\rho^{*}$ be the maximizer of $-\rho R-\ln\max\left\{ \phi\left(\rho\right),\,\Psi\right\} $,
and define 
\[
\tau_{2}\left(s\right)=\begin{cases}
\rho^{*}-\epsilon & \text{ when }s<\xi\\
\tau\left(s\right) & \text{ when }s\ge\xi,
\end{cases}
\]
and 
\[
g\left(s\right)=\exp\left(-\int_{\xi}^{s}\tau_{2}\left(r\right)r^{-1}\, dr\right),
\]
where $\epsilon$ and $\xi$ are suitable constants.

We now use Lemma~\ref{lem:smallprob} to \ref{lem:PsiLessThanOne} to prove the theorem. The main
idea is to design a function $g(s)$ and apply the Markov
inequality to $g(S_{n})$. Note that $\frac{d}{d\rho}\ln\phi\left(\rho\right)$
is continuous at $\rho=0$ and $\left.\frac{d}{d\rho}\ln\phi\left(\rho\right)\right|_{\rho=0}=-I(X;Y)$,
therefore, $-R\rho-\ln\max\left\{ \phi\left(\rho\right),\,\psi\right\} $
is positive when $\rho$ is small. If the proposed bound on the error
exponent holds, then $E\left(R\right)>0$ for any $R<I(X;Y)$, and
thus capacity can be achieved.

Let $\rho^{*}$ be the maximizer of $-\rho R-\ln\max\left(\phi\left(\rho\right),\,\Psi\right)$.
Since $\phi\left(\rho\right)$ is continuous, we may assume $\phi\left(\rho^{*}\right)\ge\Psi$.
Let $\epsilon>0$ and $\tau^{*}\left(s\right)>0$ be a nondecreasing
function such that 
\[
\Psi e^{\epsilon}\ge\psi_{s}\left(\tau^{*}\left(s\right)\right)
\]
for all $s\in(0,1)$.

By Lemma \ref{lem:smallprob}, there exists $\xi_{2}$ such that when
$s\le\xi_{2}$, we have $\psi_{s}\left(\rho\right)\le\phi\left(\rho\right)e^{\epsilon}$
for $\rho\le\rho^{*}-4\epsilon/R$. Again by Lemma \ref{lem:smallprob},
there exists $\xi\le\xi_{2}$ such that when $s\le\xi$, we have $\phi\left(\rho\right)\le\psi_{s}\left(\rho\right)e^{\epsilon}$
for $\rho\le\tau^{*}\left(\xi_{2}\right)$. Define 
\[
\tau\left(s\right)=\begin{cases}
\rho^{*}-4\epsilon/R & \text{ when }s<\xi\\
\tau^{*}\left(s\right) & \text{ when }s\ge\xi.
\end{cases}
\]
Note that 
\begin{align*}
\ln\phi\left(\tau^{*}\left(\xi\right)\right) & \le\ln\psi_{\xi}\left(\tau^{*}\left(\xi\right)\right)+\epsilon\\
 & \le\ln\Psi+2\epsilon\\
 & \le\ln\phi\left(\rho^{*}\right)+2\epsilon\\
 & <\ln\phi\left(\rho^{*}-4\epsilon/R\right).
\end{align*}
This implies that $\tau^{*}\left(\xi\right)\ge\rho^{*}-4\epsilon/R$
by the convexity of $\phi\left(\rho\right)$. Hence $\tau\left(s\right)$
is nondecreasing. Define 
\[
g\left(s\right)=\exp\left(-\int_{\xi}^{s}\tau\left(r\right)r^{-1}\, dr\right).
\]
We then consider the quantity $\mathbb{E}\left[g\left(S_{i}\right)\right]$.
Note that $g\left(s\right)$ is nonincreasing, hence {\allowdisplaybreaks
\begin{align*}
 \lefteqn{\mathbb{E}\left[\left.g\left(S_{i}\right)/g\left(S_{i-1}\right)\right|S_{i-1}=s\right]}\;\; & \\
 & =\mathbb{E}\left[\left.\exp\left(-\int_{s}^{S_{i}}\tau\left(r\right)r^{-1}\, dr\right)\right|S_{i-1}=s\right]\\
 & \le\mathbb{E}\left[\left.\exp\left(-\int_{s}^{S_{i}}\tau\left(s\right)r^{-1}\, dr\right)\right|S_{i-1}=s\right]\\
 & =\mathbb{E}\left[\left.S_{i}^{-\tau\left(s\right)}/s^{-\tau\left(s\right)}\right|S_{i-1}=s\right]\\
 & \le\max\left(\phi\left(\rho^{*}-4\epsilon/R\right)e^{\epsilon},\,\Psi e^{\epsilon}\right)\\
 & =\phi\left(\rho^{*}-4\epsilon/R\right)e^{\epsilon}.
\end{align*}
Decoding succeeds if $S_{n}\ge2/3>1/2$. Since $g\left(S_{0}\right)=e^{\left(\rho^{*}-2\epsilon/R\right)nR}/\xi^{-\left(\rho^{*}-2\epsilon/R\right)}$,
we have 
\begin{align*}
\mathbb{P}\left\{ S_{n}<2/3\right\}& \le\mathbb{E}\left[g\left(S_{n}\right)\right]/g\left(2/3\right)\\
 & \le\frac{e^{\left(\rho^{*}-4\epsilon/R\right)nR}}{\xi^{-\left(\rho^{*}-4\epsilon/R\right)}}\cdot\frac{\left(\phi\left(\rho^{*}-4\epsilon/R\right)e^{\epsilon}\right)^{n}}{g\left(2/3\right)}\\
 & =\frac{1}{\xi^{-\left(\rho^{*}-4\epsilon/R\right)}g\left(2/3\right)} \\
 &\;\;\;\cdot\exp\left(-n\!\cdot\!\left(-\rho^{*}R\!+\!\epsilon\!-\!\ln\left(\phi\left(\rho^{*}\!-\!4\epsilon/R\right)\right)\right)\right).
\end{align*}
The proof of the theorem is completed by letting $\epsilon\to0$.}

\section{Coding Complexity}\label{sec:Algorithms}

In this section, we briefly discuss the implementation of our coding algorithm and show
that its computational complexity for DMCs is
$O\left(n\log n\right)$ and its memory complexity is $O\left(n\right)$.

Although there are $e^{nR}$ possible messages, most of them share
the same pseudo posterior probability, so instead of storing the pseudo
posterior probabilities of the messages separately, we store intervals
of message points with the same pseudo posterior probability. We use
one binary search tree to keep track
of boundary points of these intervals, and another self balancing
binary search tree to keep
track of the cumulative pseudo posterior probabilities up to their
boundary points. The encoder and the decoder both keep and update
a copy of each tree (which holds the same content due to feedback).

We implemented the self balancing tree by a splay tree \cite{sleator1985}. For $n$ transmissions,
the number of nodes in the tree is at most $n\left|\mathcal{X}\right|$,
and therefore the queries and the updates can be done in $O\left(n\left|\mathcal{X}\right|\log\left(n\left|\mathcal{X}\right|\right)\right)=O\left(n\log n\right)$,
and the memory complexity is $O\left(n\right)$. Detailed description of this implementation
can be found in~\cite{feedbackcoding2013}.

\begin{figure}
\begin{centering}
\includegraphics[scale=0.5]{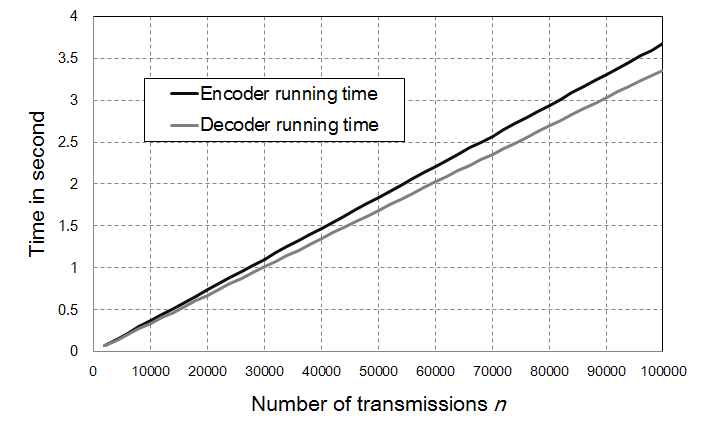}
\par\end{centering}

\begin{centering}
\includegraphics[scale=0.5]{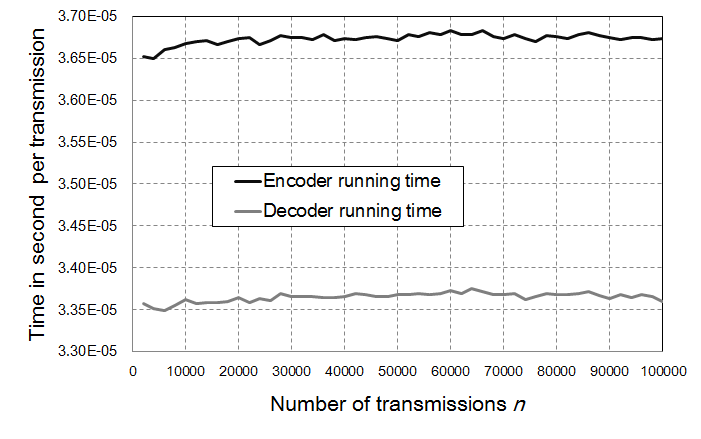}
\par\end{centering}

\begin{centering}
\includegraphics[scale=0.5]{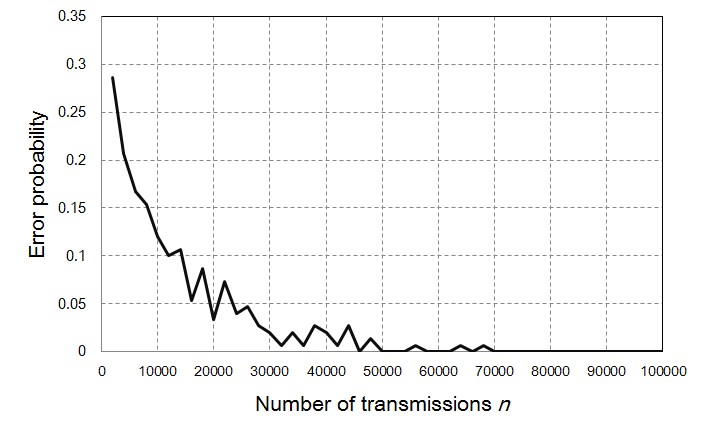} 
\par\end{centering}

\caption{\label{fig:experiment}Top: Running time of our coding algorithm for $\mathrm{BSC}(0.1)$ versus the number
of channel uses $n$. Middle: Running time divided by $n$. Bottom:
Empirical error probability (the portion of trials where the decoded
message does not match the transmitted one).}
\end{figure}

To corroborate our analysis, we performed simulations of our algorithm
assuming a $\mathrm{BSC}(0.1)$ and rate $R=0.98C$ with $n$ from
$2000$ to $100,000$. For each $n$, $150$ independent trials are
run to obtain an average running time and an estimate of the error
probability. Figure \ref{fig:experiment} shows that the average running
time is close to linear.

\section{Conclusion} \label{sec:Conclusion}

We proposed a new low coding complexity feedback coding scheme which achieves the capacity
of all DMCs. Our scheme is much easier to analyze than posterior
matching, making it possible to establish a lower bound on the error
exponent that is close to the sphere packing bound at high rate. It
would be interesting to explore if our scheme can be modified so that
the error exponent exactly coincides with the sphere packing bound
when the rate is above a certain threshold.
Another possible extension is to investigate whether our scheme achieves the channel dispersion given in~\cite{polyanskiy2010}. Although variable-length coding with feedback can achieve zero dispersion~\cite{Polyanskiy2011}, this may not be achievable using our scheme since it is fixed-length.

\section{Acknowledgments}

The authors are indebted to Young-Han Kim, Chandra Nair, Tsachy
Weissman, and the anonymous reviewers for invaluable comments that have greatly improved the exposition of the results in this paper.

\appendices

\section{Proof of Corollary~\ref{cor:ratetocapacity}}\label{coro}

Recall that the error exponent in Theorem~\ref{thm:errexp_1pass}
is 
\begin{align*}
E(R) & \ge\sup_{\rho>0}\left\{ -\rho R-\ln\max\left(\phi\left(\rho\right),\,\Psi\right)\right\} .
\end{align*}
Consider the Taylor expansion of $\ln\phi\left(\rho\right)$ at $\rho=0$,
\begin{align*}
&\ln\phi\left(\rho\right) \\
 & =\ln\sum_{y}\left(\sum_{x}p(x)p(y|x)^{1-\rho}\right)\left(\sum_{x}p(x)p(y|x)\right)^{\rho}\\
 & =\rho\cdot-I\left(X;Y\right)+\frac{\rho^{2}}{2}\cdot\sigma^{2}+O\left(\rho^{3}\right),
\end{align*}
where 
\begin{align*}
 & \sigma^{2}=\mathrm{Var}[\ln(p(Y|X)/p(Y))]\\
 & \;\;=\mathbb{E}\left[\ln(p(Y|X)/p(Y))^{2}\right]-\mathbb{E}\left[\ln(p(Y|X)/p(Y))\right]^{2}\\
 & \;\;=\!\sum_{x}\!\sum_{y}p(x)p(y|x)\!\left(\ln\frac{p(y|x)}{\sum_{x'}p(x')p(y|x')}\right)^{2}\!\!\!-\!I\left(X;\!Y\right)^{2}\!.
\end{align*}
Take $\rho=\sigma^{-2}\left(I\left(X;Y\right)-R\right)$. As $R\to I(X;Y)$,
we have $\rho\to0$, and therefore $\phi\left(\rho\right)\to1$ will
be larger than $\Psi$, and 
\begin{align*}
E(R) & \ge-\rho R-\ln\phi\left(\rho\right)\\
 & =\rho\left(I\left(X;Y\right)-R\right)-\frac{\rho^{2}}{2}\cdot\sigma^{2}-O\left(\rho^{3}\right)\\
 & =\frac{1}{2}\sigma^{-2}\left(I\left(X;Y\right)\!-\!R\right)^{2}\!-O\left(\left(I\left(X;Y\right)-R\right)^{3}\right).
\end{align*}
This completes the proof of the corollary.

\section{Proof of Lemma~\ref{lem:smallprob} (Starting Phase MGF)}\label{lem_start}

Assume $S_{0}=s\le S_{\mathrm{start}}$, then 
\begin{align*}
S_{1} & =\int_{[0,s]+U_{1}\,\mathrm{mod}\,1}dF_{W|Y}\left(w|Y_{1}\right)\\
 & =\int_{[-s/2,s/2]+U'_{1}\,\mathrm{mod}\,1}dF_{W|Y}\left(w|Y_{1}\right),
\end{align*}
where $U'_{1}=U_{1}+\left(s/2\right)\,\mathrm{mod}\,1$. Note that
$F_{X}^{-1}(w)=x$ if $F_{X}(x-1)<w\le F_{X}(x)$. Let $\alpha=s/\min_{x}p\left(x\right)$,
and let $A$ be the event that 
\begin{align*}
&\left(1-\frac{\alpha}{2}\right)F_{X}(x-1)+\frac{\alpha}{2}F_{X}(x)<U'_{1} \\
&\;\;\;\;\;\;\;\;\;\;\;\;\;\;\;\;\;\;\;\;\;\;\;\;\;\;\;\;\le\frac{\alpha}{2}F_{X}(x-1)+\left(1-\frac{\alpha}{2}\right)F_{X}(x)
\end{align*}
for some $x$. Then $\mathbb{P}\left\{ A\right\} =1-\alpha$. Note
that $A$ is independent of $F_{X}^{-1}\left(U'_{1}\right)$ (the
input symbol that $U'_{1}$ is mapped to). Conditioned on $A$, the
intervals $[-s/2,s/2]+U'_{1}\,\mathrm{mod}\,1$ does not cross the
boundary points $F_{X}(x)$, and we have $S_{1}=S_{0}+\ln p_{X|Y}\left(F_{X}^{-1}\left(U'_{1}\right)|Y_{1}\right)/p_{X}\left(F_{X}^{-1}\left(U'_{1}\right)\right)$.

We now show that $S_{0}/S_{1}$ is almost surely bounded by a constant
independent of $S_{0}=s$. Note that 
\[
S_{0}/S_{1}\ge\min_{x,y:p\left(x|y\right)>0}\frac{p\left(x\right)}{p\left(x|y\right)}\overset{def}{=}\omega_{\mathrm{lower}}
\]
almost surely. Next we establish an upper bound. If $p\left(x|y\right)>0$
for any $x,y$, then 
\[
S_{0}/S_{1}\le\max_{x,y}\frac{p\left(x\right)}{p\left(x|y\right)}.
\]
Note that when $S_{0}\le\min_{x}p\left(x\right)$, the interval $[0,s)+U_{1}\,\mathrm{mod}\,1$
intersects at most one boundary point $F_{X}(x)$. Assume $F_{X}(i)\in\left([0,s)+U_{1}\,\mathrm{mod}\,1\right)$,
and let $r=s^{-1}\left(F_{X}(i)-(U_{1}\,\mathrm{mod}\,1)\right)$
be the portion of the interval lying in the $X=i$ region, then the
maximal information gain scheme would select $X$ among $x\in\{i,i+1\}$
that gives a larger 
\[
\mathbb{E}\left[\left.\ln\left(rp\left(i|Y\right)+\left(1-r\right)p\left(i+1|Y\right)\right)\right|X=x\right]\overset{def}{=}b_{x}\left(r\right).
\]
If we have $p_{Y|X}\left(y|i+1\right)>0$ for any $y$ with $p_{Y|X}\left(y|i\right)>0$,
then $S_{0}/S_{1}\le\max_{x,y:p(x|y)>0}p(x)/p(x|y)$ holds when $X=i$.
Otherwise there exists a $y$ such that $p_{Y|X}\left(y|i\right)>0$,
$p_{Y|X}\left(y|i+1\right)=0$, then $b_{i}\left(r\right)\to-\infty$
when $r\to0$. By continuity, assume $b_{i+1}\left(r\right)>b_{i}\left(r\right)$
for $r<r_{i,i+1}$, then when $X=i$ we have $r\ge r_{i,i+1}$, and
\[
S_{0}/S_{1}\le r_{i,i+1}^{-1}\max_{x,y:p\left(x|y\right)>0}\frac{p\left(x\right)}{p\left(x|y\right)}
\]
almost surely. Define $r_{i+1,i}$ similarly. Therefore $S_{0}/S_{1}\le\omega_{\mathrm{upper}}\overset{def}{=}\big(\max_{i,j}r_{i,j}^{-1}\big)\big(\max_{x,y:p(x|y)>0}p(x)/p(x|y)\big)$,
and 
\begin{align*}
&\left(1-\alpha\right)\phi\left(\rho\right)+\alpha\omega_{\mathrm{lower}}^{\rho}\le\mathbb{E}\left[\left.S_{1}^{-\rho}/S_{0}^{-\rho}\right|S_{0}=s\right] \\
&\;\;\;\;\;\;\;\;\;\;\;\;\;\;\;\;\;\;\;\;\;\;\;\;\;\;\;\;\;\;\;\;\;\;\;\;\;\;\;\;\;\;\;\;\;\;\le\left(1-\alpha\right)\phi\left(\rho\right)+\alpha\omega_{\mathrm{upper}}^{\rho}.
\end{align*}
The proof of Lemma~\ref{lem:smallprob} is completed by letting $\omega=\max\left\{ \omega_{\mathrm{upper}},\omega_{\mathrm{lower}}^{-1}\right\} $. 

\section{Proof of Lemma~\ref{lem:largeprob} (Ending Phase MGF)}\label{lem_end}

Assume $S_{0}=s\ge1-\xi$. Note that the interval of the message $[0,s]+U_{1}\,\mathrm{mod}\,1$
overlaps all the intervals corresponding to the input symbols.
Therefore the encoder can choose among all symbols the one that minimizes
the expected value of $-\ln S_{1}$. 
\begin{align*}
S_{1} & =\int_{[0,s]+U_{1}\,\mathrm{mod}\,1}dF_{W|Y}\left(w|Y_{1}\right)\\
 & =1-\int_{[-\left(1-s\right)/2,\left(1-s\right)/2]+U'_{1}\,\mathrm{mod}\,1}dF_{W|Y}\left(w|Y_{1}\right),
\end{align*}
where $U'_{1}=U{}_{1}+\left(\left(1+s\right)/2\right)\,\mathrm{mod}\,1$.
Note that $F_{X}^{-1}(w)=x$ if $F_{X}(x-1)<w\le F_{X}(x)$. Let $\alpha=\left(1-s\right)/\min_{x}p\left(x\right)$,
and let $A$ be the event that 
\begin{align*}
&\left(1-\frac{\alpha}{2}\right)F_{X}(x-1)+\frac{\alpha}{2}F_{X}(x)<U'_{1} \\
&\;\;\;\;\le\frac{\alpha}{2}F_{X}(x-1)+\left(1-\frac{\alpha}{2}\right)F_{X}(x)
\end{align*}
for some $k$. Then $\mathbb{P}\left\{ A\right\} =1-\alpha$. Note
that $A$ is independent of $F_{X}^{-1}\left(U'_{1}\right)$ (the
input symbol that $U'_{1}$ is mapped to).

Conditioned on $A$ and $U'_{1}=u'_{1}$, the intervals $[-\left(1-s\right)/2,\left(1-s\right)/2]+U'_{1}\,\mathrm{mod}\,1$
does not cross the boundary points $F_{X}(x)$. Assume the interval
maps to $x_{1}=F_{X}^{-1}\left(u'_{1}\right)$. Define the opposite
symbol $\mathrm{opp}\left(x_{1}\right)$ as the symbol $\bar{x}_{1}$
that minimizes 
\[
\mathbb{E}\left[\left.p_{X|Y}\left(x_{1}|Y\right)\right|X=\bar{x}_{1}\right].
\]
In case of a tie, choose the symbol that minimizes $\mathbb{E}\left[\left.\left(p_{X|Y}\left(x_{1}|Y\right)\right)^{2}\right|X=\bar{x}_{1}\right]$,
and so on. Since 
\begin{align*}
 & \mathbb{E}\left[\left.-\ln S_{1}\right|U'_{1}=u'_{1},X=x\right]\\
 & =\mathbb{E}\Bigl[-\ln\Bigl(1-\int_{[-\frac{1-s}{2},\frac{1-s}{2})+u'_{1}\,\mathrm{mod}\,1}\!\!\!\!\!\!\!\!\!\!\!\!\!\!\!\!\!\!\!\!\!\!\!\!\!\!\!\!dF_{W|Y}\left(w|Y\right)\Bigr)\Bigr|X=x\Bigr]\\
 & =\mathbb{E}\Bigl[-\ln\Bigl(1-\frac{1-s}{p_{X}\left(x_{1}\right)}p_{X|Y}\left(x_{1}|Y\right)\Bigr)\Bigr|X=x\Bigr]\\
 & =\sum_{k=1}^{K}k^{-1}\Bigl(\frac{1-s}{p_{X}\left(x_{1}\right)}\Bigr)^{k}\mathbb{E}\left[\left.\left(p_{X|Y}\left(x_{1}|Y\right)\right)^{k}\right|X=x\right] \\
 &\;\;\;\;+O\Bigl(\left(1-s\right)^{K+1}\Bigr)
\end{align*}
by the Taylor series expansion, we can find $S_{\mathrm{end}}$ such
that the maximal information gain scheme chooses $\mathrm{opp}\left(x_{1}\right)=\mathrm{opp}\left(F_{X}^{-1}\left(u'_{1}\right)\right)$
whenever $s\ge S_{\mathrm{end}}$ and $u'_{1}$ satisfies the conditions
of the event $A$.

Note that $p_{X}\left(x_{1}\right)$ is the weighted mean of $\mathbb{E}\left[\left.p_{X|Y}\left(x_{1}|Y\right)\right|X=\bar{x}_{1}\right]$
over $\bar{x}_{1}$, and those values are not all equal (or else the
capacity of the channel is zero), we have, for any $x_{1}$, 
\[
\mathbb{E}\left[\left.p_{X|Y}\left(x_{1}|Y\right)\right|X=\mathrm{opp}\left(x_{1}\right)\right]\le\left(1-\eta\right)p_{X}\left(x_{1}\right)
\]
for a constant $\eta>0$ which does not depend on $x_{1}$.

Assume $S_{\mathrm{end}}$ is close enough to $1$ such that $s^{-\left(1-s\right)^{-1}}\ge e^{1-\eta/4}$
for $s\ge S_{\mathrm{end}}$. 
\begin{align*}
\lefteqn{\mathbb{E}\left[\left.S_{1}^{-\gamma\left(1-s\right)^{-1}}\right|A,S_{0}=s\right]}\\
 & =\mathbb{E}\biggl[\biggl(1-\frac{p_{X|Y}\left(F_{X}^{-1}\left(U'_{1}\right)|Y\right)}{p_{X}\left(F_{X}^{-1}\left(U'_{1}\right)\right)}\left(1-s\right)\biggr)^{-\gamma\left(1-s\right)^{-1}} \\
 &\;\;\;\;\;\;\;\;\;\;\;\; \biggl|X=\mathrm{opp}\left(F_{X}^{-1}\left(U'_{1}\right)\right),S_{0}=s\biggr]\\
 & \le\max_{x\in\mathcal{X}}\mathbb{E}\biggl[\Big(1-\frac{p_{X|Y}\left(x|Y\right)}{p_{X}\left(x\right)}\left(1-s\right)\Big)^{-\gamma\left(1-s\right)^{-1}} \\
 &\;\;\;\;\;\;\;\;\;\;\;\;\biggl|X=\mathrm{opp}\left(x\right),S_{0}=s\biggr]\\
 & \le\max_{x\in\mathcal{X}}\mathbb{E}\biggl[\exp\Bigl(\gamma\Bigl(\frac{p_{X|Y}\left(x|Y\right)}{p_{X}\left(x\right)}+\frac{\eta}{8}\Bigr)\Bigr) \\
 &\;\;\;\;\;\;\;\;\;\;\;\;\biggl|X=\mathrm{opp}\left(x\right),S_{0}=s\biggr]\\
 & \le\max_{x\in\mathcal{X}}\exp\biggl(\mathbb{E}\biggl[\gamma\Bigl(\frac{p_{X|Y}\left(x|Y\right)}{p_{X}\left(x\right)}+\frac{\eta}{4}\Bigr) \\
 &\;\;\;\;\;\;\;\;\;\;\;\;\bigg|X=\mathrm{opp}\left(x\right),S_{0}=s\biggr]\biggr)\\
 & \le\exp\left(\gamma\left(\left(1-\eta\right)+\eta/4\right)\right)\\
 & =e^{\gamma\left(1-\eta/4\right)}e^{-\eta\gamma/2}
\end{align*}
for $S_{\mathrm{end}}$ is close enough to $1$ and $\gamma=\gamma\left(\eta,S_{\mathrm{end}}\right)>0$
small enough (depend only on the channel, $\eta$ and $S_{\mathrm{end}}$).
The third line from the bottom can be shown by differentiating the
expressions with respect to $\gamma$. We have 
\begin{align*}
\lefteqn{\mathbb{E}\left[\left.S_{1}^{-\gamma\left(1-s\right)^{-1}}/S_{0}^{-\gamma\left(1-s\right)^{-1}}\right|A,S_{0}=s\right]}\qquad&\\
 & \le e^{-\gamma\left(1-\eta/4\right)}\mathbb{E}\left[\left.S_{1}^{-\gamma\left(1-s\right)^{-1}}\right|A,S_{0}=s\right]\\
 & \le e^{-\eta\gamma/2}.
\end{align*}
Define $\omega=\max_{x,y:p\left(x\right)>0}\frac{p\left(x|y\right)}{p\left(x\right)}$,
then 
\begin{align*}
S_{1}^{-\gamma\left(1-s\right)^{-1}}/S_{0}^{-\gamma\left(1-s\right)^{-1}} & \le\left(1-\omega\left(1-s\right)\right)^{-\gamma\left(1-s\right)^{-1}}s^{\gamma\left(1-s\right)^{-1}}\\
 & \le e^{\gamma\left(1+\omega-\eta/4\right)}.
\end{align*}
Hence, 
\begin{align*}
 \psi_{s}\left(\gamma\left(1-s\right)^{-1}\right)& =\left(1-\alpha\right)e^{-\eta\gamma/2}+\alpha e^{\gamma\left(1+\omega-\eta/4\right)}\\
 & \le e^{-\eta\gamma/2}+\frac{1-s}{\min_{x_{k}}p\left(x_{k}\right)}e^{\gamma\left(1+\omega-\eta/4\right)}\\
 & \le e^{-\eta\gamma/4}
\end{align*}
for $1-s$ small enough. This completes the proof of Lemma~\ref{lem:largeprob}.

\section{Proof of Lemma~\ref{lem:PsiLessThanOne}}\label{lem_trans}

By Lemma \ref{lem:largeprob}, when $S_{i}\ge S_{\mathrm{end}}$,
the actual moment generating function can be bounded. It is left to
bound the actual MGF for $S_{i}<S_{\mathrm{end}}$. We first prove
that $\psi_{s}'\left(0\right)$ for $s\le S_{\mathrm{end}}$ can be
bounded above and away from $0$.

Since the maximal information gain rule~\eqref{eq:maxinfogain}
minimizes the expectation of $-\ln S_{i}$, it has a smaller $\mathbb{E}\left[-\ln S_{i}\right|S_{i-1}=s]$
than any other rule of selecting $V_{i}$. In particular, if we generate
$V_{i}$ according to $\mathrm{U}[0,1]$, the expectation would be
$\tilde{\mathbb{E}}[-\ln S_{i}|S_{i-1}=s]$, where $\tilde{\mathbb{E}}$
denotes the expectation under the probability measure $\tilde{\mathbb{P}}$.
Therefore, {\allowdisplaybreaks 
\begin{align*}
\lefteqn{\mathbb{E}[-\ln S_{i}|S_{i-1}=s]}\\
 & \le\tilde{\mathbb{E}}[-\ln S_{i}|S_{i-1}=s]\\
 & =\tilde{\mathbb{E}}\biggl[-\ln\int_{[T_{i-1},T_{i-1}+s]+U_{i}\,\mathrm{mod}\,1} f_{W|Y}(w|Y_i) \, dw\biggr]\\
 & =\tilde{\mathbb{E}}\biggl[-\ln\int_{[T_{i-1},T_{i-1}+s]+U_{i}\,\mathrm{mod}\,1}\frac{f_{Y|W}(Y_{i}|w)}{f_{Y}(Y_{i})}\, dw\biggr]\\
 & =\tilde{\mathbb{E}}\biggl[-\ln\int_{[0,s]+U_{i}\,\mathrm{mod}\,1}\frac{f_{Y|W}(Y_{i}|w)}{f_{Y}(Y_{i})}\, dw \,\Big|\, T_{i-1}=0 \biggr]\\
 & =\int_{0}^{1}\!\int_{[0,s]+u\,\mathrm{mod}\,1}\int\!\biggl(\!-\!\ln\!\int_{[0,s]+u\,\mathrm{mod}\,1}\!\!\!\frac{f_{Y|W}(y|w)}{f_{Y}(y)}\, dw\!\biggr) \\
 &\;\;\;\cdot f_{Y|W}\left(y|w_{0}\right)dy\cdot s^{-1}\, dw_{0}\cdot du\\
 & =-\ln s+\int_{0}^{1}\int\biggl(\int_{[0,s]+u\,\mathrm{mod}\,1}f_{Y|W}\left(y|w\right)\cdot s^{-1}\, dw\biggr) \\
 &\;\;\;\;\biggl(-\ln\frac{\int_{[0,s]+u\,\mathrm{mod}\,1}f_{Y|W}(y|w)\cdot s^{-1}\, dw}{f_{Y}(y)}\biggr)dy\cdot du\\
 & =-\ln s-I(U_{i};Y_{i}|M,S_{i-1}=s).
\end{align*}
}Hence 
\begin{align*}
\psi_{s}'\left(0\right) & =\left.\frac{d}{d\rho}\psi_{s}\left(\rho\right)\right|_{\rho=0}\\
 & =\mathbb{E}[-\ln S_{i}/s|S_{i-1}=s]\\
 & \le-I(U_{i};Y_{i}|M,S_{i-1}=s)\\
 & =-H(Y_{i})+H(Y_{i}|U_{i},M=m,S_{i-1}=s).
\end{align*}

Since $H(Y_{i}|U_{i}=u,M,S_{i-1}=s)$ is continuous in $u$, and entropy
is strictly concave, to show $H(Y_{i}|U_{i},M,S_{i-1}=s)<H(Y_{i})$,
it suffices to show that $Y_{i}$ does not have the same distribution
conditioned on $U_{i}=u$ for different $u$. Assume the contrary,
i.e., that there exists some $s<1$ such that $Y_{i}$ has the same
distribution conditioned on $U_{i}=u$ and $S_{i-1}=s$ for all $u$.
Note that if $V_{i}\sim\mathrm{U}[0,1]$, 
\[
\mathbb{P}\left\{ Y_{i}=y|U_{i}=u\right\} =s^{-1}\int_{[0,s]+u\,\mathrm{mod}\,1}p_{Y|W}\left(y|w\right)\, dw.
\]
Differentiating the expression with respect to $u$, we have 
\[
p_{Y|W}\left(y|w\right)=p_{Y|W}\left(y|w+s\,\,\mathrm{mod}\,1\right)
\]
for all $y$ and $w$. By $p_{Y|W}(y|w)=p_{Y|X}(y|F_{X}^{-1}(w))$
and the assumption that the channel has no redundant input symbols,
we have 
\[
F_{X}^{-1}(w)=F_{X}^{-1}(w+s\,\,\mathrm{mod}\,1)
\]
for all $y$ and $w$. This implies $F_{X}^{-1}(w)$ is either constant
or periodic, which leads to a contradiction since $F_{X}^{-1}(w)$
is nondecreasing and is not constant. Therefore we know that $H(Y_{i}|U_{i},M,S_{i-1}=s)<H(Y_{i})$
for $s<1$. Since $H(Y_{i}|U_{i},M,S_{i-1}=s)$ is continuous in $s\in[0,S_{\mathrm{end}}]$
assuming $V_{i}\sim\mathrm{U}[0,1]$, the expression is bounded above
and away from $H(Y_{i})$, and thus we have $\psi_{s}'\left(0\right)\le\zeta$
for all $s\le S_{\mathrm{end}}$, where $\zeta<0$ is a constant.

Without loss of generality, assume the message transmitted is $m=1$,
then the message interval at time $i$ is $[U_{i},U_{i}+S_{i-1}]\mod1$,
and the symbol selected by the maximal information gain scheme is
a function $X_{i}=x^{*}\left(S_{i-1},U_{i}\right)$ of $S_{i-1}$
and $U_{i}$. Therefore 
\begin{align*}
&\psi_{s}\left(\rho\right) \\ &=\mathbb{E}\left[\left.S_{i}^{-\rho}/S_{i-1}^{-\rho}\right|S_{i-1}=s\right]\\
 & =\int_{0}^{1}\mathbb{E}\left[\left.S_{i}^{-\rho}/S_{i-1}^{-\rho}\right|S_{i-1}=s,U_{i}=u,X_{i}=x^{*}\left(s,u\right)\right]\! du\\
 & =\int_{0}^{1}\psi_{s,u,x^{*}\left(s,u\right)}\left(\rho\right)\, du,
\end{align*}
where $\psi_{s,u,x}\left(\rho\right)=\mathbb{E}\!\left[\left.S_{i}^{-\rho}/S_{i-1}^{-\rho}\right|S_{i-1}\!=\!s,U_{i}\!=\!u,X_{i}\!=\!x\right]$
is the moment generating function when the message interval is $[u,u+s]\mod1$
and the transmitted symbol is $x$.

It is easy to show that $\psi'_{s,u,x}\left(\rho\right)$, when treated
as a function of $\left(s,u,\rho\right)$, is continuous and strictly
increasing in $\rho$. Restricted on $s\le S_{\mathrm{end}}$ and
$\rho\le1$, the domain of the function is $[0,S_{\mathrm{end}}]\times[0,1]\times[0,1]$
which is compact, and therefore the function is uniformly continuous
in this domain. We can find $\bar{\rho}_{x}>0$ such that $\psi'_{s,u,x}\left(\rho\right)-\psi'_{s,u,x}\left(0\right)\le-\zeta/2$
for any $s\le S_{\mathrm{end}}$, $u\in[0,1]$ and $\rho\le\bar{\rho}_{x}$.
Let $\bar{\rho}=\min_{x}\bar{\rho}_{x}$. For any $s\le S_{\mathrm{end}}$
and $\rho\le\bar{\rho}$, 
\begin{align*}
\psi_{s,u,x}\left(\rho\right) & =1+\int_{0}^{\rho}\psi'_{s,u,x}\left(r\right)\, dr\\
 & \le1+\rho\left(\psi'_{s,u,x}\left(0\right)-\zeta/2\right),
\end{align*}
and 
\begin{align*}
\psi_{s}\left(\rho\right) & =\int_{0}^{1}\psi_{s,u,x^{*}\left(s,u\right)}\left(\rho\right)\, du\\
 & \le\int_{0}^{1}\left(1+\rho\left(\psi'_{s,u,x^{*}\left(s,u\right)}\left(0\right)-\zeta/2\right)\right)du\\
 & =1+\rho\left(\psi'_{s}\left(0\right)-\zeta/2\right)\\
 & \le1+\rho\zeta/2.
\end{align*}
Let 
\[
\tau\left(s\right)=\begin{cases}
\min\left(\bar{\rho},\,\gamma\left(1-S_{\mathrm{end}}\right)^{-1}\right) & \text{ when }s<S_{\mathrm{end}}\\
\gamma\left(1-s\right)^{-1} & \text{ when }s\ge S_{\mathrm{end}}
\end{cases}
\]
be a nondecreasing function, where $\gamma$ is from Lemma \ref{lem:largeprob}.
Then 
\begin{align*}
\Psi & \le\sup_{s\in(0,1)}\psi_{s}\left(\tau\left(s\right)\right)\\
 & \le\max\left(1+\min\left(\bar{\rho},\,\gamma\left(1-S_{\mathrm{end}}\right)^{-1}\right)\cdot\zeta/2,\,\Psi_{\mathrm{end}}\right)\\
 & <1.
\end{align*}
This completest the proof of Lemma~\ref{lem:PsiLessThanOne}.

\bibliographystyle{IEEEtran}
\bibliography{ref,nit}

\end{document}